\documentclass[a4paper,aps,prd,10pt,preprintnumbers,showpacs,twocolumn,superscriptaddress,nofootinbib,amsmath,amssymb]{revtex4-1}
\usepackage{cmap}
\usepackage[utf8]{inputenc}
\usepackage[T1]{fontenc}

\def\Order#1{{\cal O}\left(#1\right)}

\begin{document}

\title{Dymnikova black hole from an infinite tower of higher-curvature corrections}
\author{R. A. Konoplya}
\email{roman.konoplya@gmail.com}
\affiliation{Research Centre for Theoretical Physics and Astrophysics, Institute of Physics, Silesian University in Opava, Bezručovo náměstí 13, CZ-74601 Opava, Czech Republic}

\author{A. Zhidenko}
\email{olexandr.zhydenko@ufabc.edu.br}
\affiliation{Centro de Matemática, Computação e Cognição (CMCC), Universidade Federal do ABC (UFABC), \\ Rua Abolição, CEP: 09210-180, Santo André, SP, Brazil}

\begin{abstract}
Recently, in \cite{Bueno:2024dgm}, it was demonstrated that various regular black hole metrics can be derived within a theory featuring an infinite number of higher curvature corrections to General Relativity. Moreover, truncating this infinite series at the first few orders already yields a reliable approximation of the observable characteristics of such black holes \cite{Konoplya:2024hfg}. Here, we further establish the existence of another regular black hole solution, particularly the $D$-dimensional extension of the Dymnikova black hole, within the equations of motion incorporating an infinite tower of higher-curvature corrections. This solution is essentially nonperturbative in the coupling parameter, rendering the action, if it exists, incapable of being approximated by a finite number of powers of the curvature. In addition, we compute the dominant quasinormal frequencies of such black holes using both the Bernstein polynomial method and the 13th order WKB method with Padé approximants, obtaining a high degree of agreement between them.
\end{abstract}

\maketitle

\section{Introduction}
The singularity problem for the classical Schwarzschild solution marks the limits of General Relativity. While a great number of approaches to solve this problem were suggested recently a large class of regular black holes in $D \geq 5$ spacetime dimensions have been obtained as a result of adding the infinite number of higher-curvature terms to the Einstein action \cite{Bueno:2024dgm}. This approach is a part of the quasi-topological gravity \cite{Oliva:2010eb,Myers:2010ru,Dehghani:2011vu,Ahmed:2017jod,Cisterna:2017umf}, so that no exotic or other poorly motivated matter is introduced to produce a regular black hole. This approach has been recently studied in \cite{DiFilippo:2024mwm} and quasinormal modes of regular black holes obtained within this approach have been analyzed in \cite{Konoplya:2024hfg}.

An observation done in \cite{Konoplya:2024hfg} consisted in the justification of the lower order expansions which could serve as good approximations for observable quantities for the full regular black hole solutions. The latter is important, because,  motivated by the low-energy limit of string theory, a great area of research is devoted to various theories with higher curvature corrections, which are truncation of the initially infinite series, producing, usually, second, third or higher order theories, such as Einstein-Gauss-Bonnet or  Einstein-Lovelock ones \cite{Boulware:1985wk,Lovelock:1971yv,Myers:1988ze}.

The freedom to establish various relations between the coupling parameters allowed to produce classes of regular black-hole metrics with desired features \cite{DiFilippo:2024mwm}. The natural question is then whether it is indeed so, and any regular black hole metric can be a solution to some theory which could represented as an infinite series of higher curvature corrections and, first of all, whether such a theory corresponds to some particular choice of the coupling parameters of the quasi-topological gravity?

For this purpose we will study what kind of higher curvature corrections correspond to the $D$-dimensional generalization of the Dymnikova regular black hole metric \cite{Dymnikova:1992ux}. Dymanikova black hole can be interpreted as a solution of the Einstein equations (four or higher dimensional) with an energy-momentum tensor, corresponding to the exponentially decaying matter \cite{Dymnikova:1992ux,Paul:2023pqn} or as a solution of the Lovelock theory with a matter source \cite{Estrada:2019qsu}.

We will also delve into the fundamental characteristics of the $D$-dimensional Dymnikova black hole geometry, particularly its dominant quasinormal frequencies. These frequencies represent the inherent oscillations of black holes, determined solely by the black hole parameters and unaffected by the perturbation method \cite{Konoplya:2011qq}.
The literature on quasinormal modes of black holes in theories with higher curvature corrections, as well as various regular black holes, is vast. Typically, these theories involve perturbative approaches in coupling theories (see, for recent examples, \cite{Dubinsky:2024aeu,Konoplya:2023ahd,Konoplya:2020bxa,Zinhailo:2023xdz,Konoplya:2022hll} and references therein).
The issue of quasinormal modes for scalar fields in the above $D$-dimensional Dymnikova regular black holes has recently been addressed in \cite{Macedo:2024dqb}. However, it cannot be deemed a complete resolution, as only two frequencies were demonstrated for specific parameters of the black hole at $D=5$ in Fig.~4, without providing insights into how these modes vary with the black hole parameters. Furthermore, the use of the WKB method, which converges only asymptotically, raises concerns about the precision of the results obtained.

The paper is structured as follows: In Section~\ref{sec:HDDymnikova}, we introduce the $D$-dimensional Dymnikova black hole within the framework of an infinite series of higher curvature corrections and demonstrate its incompatibility as a solution of such a perturbative theory. In Section~\ref{sec:discussion}, we propose an analogy to elucidate the concept that a non-perturbative theory may offer a solution to such a regular black hole. Section~\ref{sec:QNMs} is dedicated to calculating the quasinormal modes of these black holes. In Section~\ref{sec:stability} we show that the scalar and gravitational fields are stable against small perturbations in the higher-dimensional Dymnikova background. Finally, the Conclusions section provides a summary of the obtained results.

\section{Generalized Dymnikova solution in pure gravity}\label{sec:HDDymnikova}
Bueno, Cano, and Hennigar~\cite{Bueno:2024dgm} considered the action of the Einsteinian theory with general higher curvature corrections
\begin{equation}\label{QTaction}
I_{\rm QT}=\frac{1}{16\pi G} \int \mathrm{d}^Dx \sqrt{|g|} \left[R+\sum_{n=2}^{n_{\rm max}} \alpha_n \mathcal{Z}_n \right]\, ,
\end{equation}
where $\alpha_n$ are arbitrary coupling constants with dimensions of length$^{2(n-1)}$ and $\mathcal{Z}_n$ are the Quasi-topological densities \cite{Bueno:2019ycr}. Regular black-hole solutions correspond to theories with the infinite number of terms in the above action.

The line element for the spherically symmetric $D$-dimensional black hole is
\begin{eqnarray}\label{metric}
ds^2 &=& - f(r) dt^2 + \frac{dr^2}{f(r)} + r^2 d\Omega^2_{D-2} \, ,
\\\nonumber &&f(r)\equiv1-r^2\psi(r)\, .
\end{eqnarray}
The equations of motion imply that
\begin{equation}\label{eqmotion}
\dfrac{d}{dr}\left(r^{D-1}h(\psi(r))\right)=0,
\end{equation}
where the function $h(\psi)$ is given by the series,
\begin{equation}\label{hseries}
h(\psi) \equiv \psi + \sum_{n=2}^{n_{\rm max}} \alpha_n \psi^n\, .
\end{equation}

The solution to Eq.~\ref{eqmotion} is
\begin{equation}\label{hequation}
h(\psi(r)) = \dfrac{\mu}{r^{D-1}}\, ,
\end{equation}
where $\mu$ is the positive constant proportional to the ADM mass.

In~\cite{Bueno:2024dgm} a limit of indefinite tower of higher curvature corrections in (\ref{QTaction}) is introduced via an explicit function $h(\psi)$, which is monotonously growing for $0<\psi<\psi_0$ and diverges at $\psi=\psi_0$. In this way, the solution to the equation (\ref{hequation}) gives a black-hole configuration, in which, by construction, the singularity at $r=0$ is replaced by the de Sitter core,
\begin{equation}\label{origin}
f(r)=1+\Order{r^2}\approx1-r^2\psi_0.
\end{equation}

Since the function $h(\psi)$ is initially defined through the series (\ref{hseries}), it is implicitly assumed that $h(\psi)$ is regular in $\psi=0$. We note that regularity in this point is not a necessary condition. The reason for that is that $\psi=0$ corresponds to the asymptotic infinity of the spacetime, which is also not a regular point. Moreover, in the present paper we propose a simple example of such a function $h(\psi)$, which is not regular in $\psi=0$, while leading to a meaningful black-hole configuration free from singularities.

A simple function of $\psi\geq0$, which is not regular at $\psi=0$, is $e^{-1/\alpha\psi}$, where $\alpha$ is a positive constant. This is a monotonous analytic function for $\psi>0$, which cannot be expanded in Taylor series near the origin. Keeping in mind this feature, we are in a position to construct the function $h(\psi)$, which also diverges at $\psi=\psi_0\equiv\alpha^{-1}>0$. We define the function as follows:
\begin{equation}\label{hform}
h(\psi)\equiv\dfrac{\psi}{1+\alpha\psi W_0\left(-\dfrac{e^{-1/\alpha\psi}}{\alpha\psi}\right)}, \quad \alpha>0,
\end{equation}
where $W_0(z)$ is the principal Lambert function, which satisfies the equation
$$W_0(z)e^{W_0(z)}=z,$$
and
$$W_0(0)=0,\quad W_0(-e^{-1})=-1.$$

For the function $h(\psi)$ given by (\ref{hform}) the solution to the Eq.~(\ref{hequation}) can be found in an explicit form,
\begin{equation}\label{psi}
\psi(r)=\dfrac{\mu}{r^{D-1}}\left(1-e^{-\dfrac{r^{D-1}}{\alpha\mu}}\right),
\end{equation}
and the corresponding metric function is
\begin{equation}\label{HDDymnikova}
f(r)=1-\dfrac{\mu}{r^{D-3}}\left(1-e^{-\dfrac{r^{D-1}}{\alpha\mu}}\right).
\end{equation}

The metric (\ref{HDDymnikova}) is regular for $r\geq0$. It coincides with the one obtained in higher dimensional Einstein theory in the presence of matter \cite{Paul:2023pqn} and, when $D=4$, it is reduced to the well-known Dymnikova black hole \cite{Dymnikova:1992ux,Platania:2019kyx}.

It is important to stress out that $h(\psi)$ cannot be represented as series (\ref{hseries}), and consequently, the action cannot be represented as (\ref{QTaction}).
Although the limit of $\alpha\to0$ clearly exists and is the Schwarzschild solution to General Relativity, the theory is not perturbative in terms of the quasi-topological corrections. Consequently, one is unable to approximate such a theory with a finite number of the higher-order corrections.

\section{Discussion}\label{sec:discussion}
Although we are unable to represent the action of the theory in the form (\ref{hseries}), it does not mean that such a theory does not exist. In order to illustrate this point we can propose an ad hoc action with a similar feature
\begin{equation}\label{adhocaction}
I=\frac{1}{16\pi G} \int \mathrm{d}^Dx \sqrt{|g|} \left(R+\dfrac{1}{\alpha}e^{-\dfrac{1}{\alpha^2R^2}}\right)\, .
\end{equation}
This theory is nonperturbative in terms of the coupling constant $\alpha$ and cannot be approximated by finite number of the higher-order curvature corrections. At the same time, it has well-defined equations of motion and approaches General Relativity in the limit of $\alpha\to0$. Thus, we cannot exclude that there is a well-defined theory of gravity, in which a $D$-dimensional Dymnikova black hole is a vacuum solution. We conclude that this theory must be nonperturbative in terms of the higher-order curvature corrections.
A similar feature has been recently considered in \cite{Koshelev:2024wfk}, where the absence of the singularity in the spherically symmetric black hole is governed by the exponential form for the graviton propagator in a ghost-free infinite derivative gravity, leading to the nonperturbative total energy of the black hole with respect to the regularization parameter.

\section{Quasinormal modes}\label{sec:QNMs}

The general relativistic equations for a scalar ($\Phi$) and electromagnetic ($A_\mu$) fields can be written as follows:
\begin{subequations}\label{coveqs}
\begin{eqnarray}\label{KGg}
\frac{1}{\sqrt{-g}}\partial_\mu \left(\sqrt{-g}g^{\mu \nu}\partial_\nu\Phi\right)&=&0,
\\\label{EmagEq}
\frac{1}{\sqrt{-g}}\partial_{\mu} \left(F_{\rho\sigma}g^{\rho \nu}g^{\sigma \mu}\sqrt{-g}\right)&=&0,
\end{eqnarray}
\end{subequations}
where $F_{\mu\nu}=\partial_\mu A_\nu-\partial_\nu A_\mu$ is the electromagnetic tensor.

After separation of variables, equations for the test massless fields in the background of the spherically symmetric black hole can be reduced to the wavelike form
\begin{equation}\label{wavelike}
\frac{d^2\Psi}{dr_*^2}+(\omega^2-V(r_*))\Psi(r_*)=0,
\end{equation}
where the tortoise coordinate is defined as follows:
\begin{equation}\label{tortoise}
dr_*=\frac{dr}{f(r)}.
\end{equation}

The effective potentials for the scalar ($V_0$) and electromagnetic ($V_1$ and $V_2$) \cite{Crispino:2000jx,Lopez-Ortega:2006vjp} fields have the form
\begin{subequations}\label{potentials}
\begin{eqnarray}
V_0(r) &=& f(r)\Biggl(\frac{\ell(\ell+D-3)}{r^2} \\\nonumber&&+ \frac{(D-2)(D-4)}{4r^2}f(r) + \frac{D-2}{2r}\frac{df}{dr}\Biggr);
\\
V_1(r) &=& f(r)\Biggl(\frac{\ell(\ell+D-3)}{r^2} \\\nonumber&&+ \frac{(D-2)(D-4)}{4r^2}f(r) - \frac{D-4}{2r}\frac{df}{dr}\Biggr);
\\
V_2(r) &=& f(r)\Biggl(\frac{(\ell+1)(\ell+D-4)}{r^2} \\\nonumber&&+ \frac{(D-4)(D-6)}{4r^2}f(r) + \frac{D-4}{2r}\frac{df}{dr}\Biggr).
\end{eqnarray}
\end{subequations}
Here $\ell$ is a nonnegative integer multipole number representing the decoupled angular variables. For the electromagnetic field $\ell=0$ corresponds to the nondynamical degrees of freedom \cite{Crispino:2000jx}, so that we consider only $\ell>0$.

\begin{table}
\begin{tabular}{l@{\hspace{0.5em}}|@{\hspace{0.5em}}c@{\hspace{0.5em}}|@{\hspace{0.5em}}c@{\hspace{1em}}c}
\hline
\hline
$\alpha$ & $\ell=0$ & $\ell=1$ &  WKB ($\ell=1$)\\
\hline
\multicolumn{4}{c}{$D=5$}\\
\hline
 0    & 0.53384 - 0.38338i & 1.01602 - 0.36233i & 1.016 - 0.362i \\
 0.2  & 0.53219 - 0.38038i & 1.01399 - 0.36146i & 1.012 - 0.361i \\
 0.25 & 0.52953 - 0.37503i & 1.01090 - 0.35924i & 1.009 - 0.359i \\
 0.3  & 0.52424 - 0.36680i & 1.00630 - 0.35554i & 1.004 - 0.356i \\
 0.35 & 0.51397 - 0.35818i & 1.00029 - 0.35092i & 0.998 - 0.350i \\
 0.4  & 0.50276 - 0.35550i & 0.99345 - 0.34609i & 0.991 - 0.344i \\
\hline
\multicolumn{4}{c}{$D=6$}\\
\hline
 0    & 0.88944 - 0.53310i & 1.44651 - 0.50927i & 1.447 - 0.509i \\
 0.2  & 0.88729 - 0.53039i & 1.44423 - 0.50844i & 1.443 - 0.508i \\
 0.25 & 0.88402 - 0.52546i & 1.44074 - 0.50623i & 1.439 - 0.506i \\
 0.3  & 0.87856 - 0.51785i & 1.43562 - 0.50245i & 1.426 - 0.497i \\
 0.35 & 0.86986 - 0.50870i & 1.42906 - 0.49753i & 1.426 - 0.497i \\
 0.4  & 0.85754 - 0.50122i & 1.42141 - 0.49217i & 1.418 - 0.490i \\
 0.45 & 0.84618 - 0.49784i & 1.41336 - 0.48678i & 1.409 - 0.479i \\
\hline
\multicolumn{4}{c}{$D=7$}\\
\hline
 0    & 1.27054 - 0.66578 i & 1.88140 - 0.64108i & 1.881 - 0.641i \\
 0.2  & 1.26804 - 0.66333 i & 1.87892 - 0.64033i & 1.878 - 0.641i \\
 0.25 & 1.26428 - 0.65872 i & 1.87508 - 0.63819i & 1.873 - 0.637i \\
 0.3  & 1.25838 - 0.65154 i & 1.86948 - 0.63445i & 1.866 - 0.635i \\
 0.35 & 1.24980 - 0.64268 i & 1.86236 - 0.62950i & 1.858 - 0.630i \\
 0.4  & 1.23823 - 0.63406 i & 1.85408 - 0.62396i & 1.849 - 0.620i \\
 0.45 & 1.22526 - 0.62827 i & 1.84524 - 0.61836i & 1.839 - 0.610i \\
 0.5  & 1.21436 - 0.62439 i & 1.83638 - 0.61273i & 1.828 - 0.600i \\
\hline
\hline
\end{tabular}
\caption{Quasinormal modes of the scalar field calculated using the Bernstein polynomial method and the 13th order WKB method with the Padé approximant $\widetilde{m}=7$, $\widetilde{n}=6$ for $\ell=1$, $\mu=1$}.\label{tabl:scalar}
\end{table}

\begin{table}
\begin{tabular}{l@{\hspace{2em}}|@{\hspace{1em}}c@{\hspace{1em}}|@{\hspace{1em}}c}
\hline
\hline
$\alpha$ & $V_1$ & $V_2$ \\
\hline
\multicolumn{3}{c}{$D=5$}\\
\hline
 0    & 0.75541 - 0.31546i & 0.64045 - 0.34752i \\
 0.2  & 0.75472 - 0.31316i & 0.64008 - 0.34472i \\
 0.25 & 0.75404 - 0.30894i & 0.63988 - 0.33971i \\
 0.3  & 0.75291 - 0.30207i & 0.63924 - 0.33155i \\
 0.35 & 0.75043 - 0.29287i & 0.63659 - 0.32059i \\
 0.4  & 0.74590 - 0.28296i & 0.63084 - 0.30969i \\
\hline
\multicolumn{3}{c}{$D=6$}\\
\hline
 0.   & 1.03139 - 0.44164i & 0.99214 - 0.50150i \\
 0.2  & 1.03065 - 0.43914i & 0.99123 - 0.49871i \\
 0.25 & 1.02995 - 0.43462i & 0.99024 - 0.49372i \\
 0.3  & 1.02891 - 0.42734i & 0.98870 - 0.48575i \\
 0.35 & 1.02693 - 0.41749i & 0.98583 - 0.47513i \\
 0.4  & 1.02303 - 0.40591i & 0.98057 - 0.46309i \\
 0.45 & 1.01708 - 0.39452i & 0.97329 - 0.45195i \\
\hline
\multicolumn{3}{c}{$D=7$}\\
\hline
 0.   & 1.35049 - 0.57997i & 1.36657 - 0.63915i \\
 0.2  & 1.34940 - 0.57735i & 1.36519 - 0.63647i \\
 0.25 & 1.34813 - 0.57258i & 1.36348 - 0.63158i \\
 0.3  & 1.34630 - 0.56496i & 1.36097 - 0.62383i \\
 0.35 & 1.34345 - 0.55477i & 1.35721 - 0.61359i \\
 0.4  & 1.33883 - 0.54276i & 1.35155 - 0.60180i \\
 0.45 & 1.33204 - 0.53047i & 1.34385 - 0.59009i \\
 0.5  & 1.32415 - 0.51928i & 1.33543 - 0.57956i \\
\hline
\hline
\end{tabular}
\caption{Quasinormal modes  of the electromagnetic field for $\ell=1$, $\mu=1$, calculated using the Bernstein polynomial method.}\label{tabl:EM}
\end{table}

Quasinormal modes are complex eigenvalues $\omega$ of wavelike equation~(\ref{wavelike}) corresponding to the purely ingoing wave at the event horizon and purely outgoing one at infinity. Here we use two alternative methods for finding quasinormal modes: the spectral method \cite{Jansen:2017oag} with the basis of Bernstein polynomials \cite{Fortuna:2020obg,Konoplya:2022zav} and the 13th order WKB method with Padé approximants \cite{Schutz:1985km,Iyer:1986np,Konoplya:2003ii,Matyjasek:2017psv,Konoplya:2019hlu}. As the WKB method is related in details in a review \cite{Konoplya:2019hlu} and effectively applied in a great number of works (see \cite{Bolokhov:2023ruj,Skvortsova:2023zmj,Skvortsova:2024atk,Paul:2023eep,Zinhailo:2018ska,Dubinsky:2024hmn} for recent examples), we will discuss only the Bernstein polynomial method here.

The wavelike equation~(\ref{wavelike}) always has a regular singularity at the event horizon $r=r_0$ and the irregular singularity at spatial infinity $r=\infty$. We introduce the new function,
\begin{equation}\label{reg}
\Psi(r)=e^{i\omega r}\left(1-\frac{r_0}{r}\right)^{-i\omega/f'(r_0)}y(r),
\end{equation}
so that $y(r)$ is regular for $r_0\leq r<\infty$ when the quasinormal boundary conditions are fulfilled.

Following \cite{Fortuna:2020obg}, we introduce the compact coordinate
$$u\equiv\frac{r_0}{r},$$
and represent $y(u)$ as a sum,
\begin{equation}\label{Bernsteinsum}
y(u)=\sum_{k=0}^NC_kB_k^N(u),
\end{equation}
where
$$B_k^N(u)\equiv\frac{N!}{k!(N-k)!}u^k(1-u)^{N-k}$$
are the Bernstein polynomials.

Substituting (\ref{reg}) into (\ref{wavelike}) and using a Chebyschev collocation grid of $N+1$ points, we obtain a set of linear equations with respect to $C_k$, which has nontrivial solutions iff the corresponding coefficient matrix is singular. The problem is reduced to the eigenvalue problem of a matrix pencil with respect to $\omega$, which can be solved numerically. Once the eigenvalue problem is solved, one can calculate the corresponding coefficients $C_k$ and explicitly determine the polynomial (\ref{Bernsteinsum}), which approximates the solution to the wave equation \cite{Fortuna:2020obg}.

In order to exclude the spurious eigenvalues, which appear due to finiteness of the polynomial basis in (\ref{Bernsteinsum}), we compare both the eigenfrequencies and corresponding approximating polynomials for different values of $N$. Namely, for the coinciding eigenfrequencies, $\omega^{(1)}$ and $\omega^{(2)}$, obtained, respectively, for $N=N^{(1)}$ and $N=N^{(2)}$, we calculate
$$1-\frac{|\langle y^{(1)}\;|\;y^{(2)} \rangle|^2}{||y^{(1)}||^2||y^{(2)}||^2}=\sin^2\alpha,$$
where $\alpha$ is the angle between the vectors $y^{(1)}$ and $y^{(2)}$ in the $L^2$-space. If $\alpha$ is sufficiently small, then the difference between $\omega^{(1)}$ and $\omega^{(2)}$ provides the error estimation\footnote{The Wolfram Mathematica\textregistered{} package with the implementation of the Bernstein spectral method \cite{Konoplya:2022zav} is publicly available from \url{https://arxiv.org/src/2211.02997/anc}.}. In order to obtain the dominant quasinormal mode we use $N^{(1)}=30$ and $N^{(2)}=50$.

It was demonstrated in \cite{Konoplya:2023aph} that the Bernstein polynomial method yields sufficient accuracy for calculating dominant quasinormal modes. Meanwhile, the WKB method utilized in \cite{Macedo:2024dqb} does not enable accurate determination of these modes. Here, we present, for the first time, the low-lying modes of the test scalar and electromagnetic fields, $\ell=0$ and $\ell=1$, respectively, for the higher-dimensional Dymnikova black hole. We employ the WKB method to compare the modes of the scalar field for $\ell=1$, where the WKB method offers a reliable approximation. From Tables \ref{tabl:scalar} and \ref{tabl:EM}, it can be observed that, similar to the four-dimensional black hole \cite{Konoplya:2023aph}, both the real and imaginary parts decrease as the coupling parameter increases. It's noteworthy that $\alpha=0$ corresponds to the Tangherlini black hole \cite{Tangherlini:1963bw}, for which accurate values of the quasinormal modes of the scalar field were obtained in \cite{Zhidenko:2006rs}. Moreover, it appears that the electromagnetic modes for $\ell=1$, $n=0$, of the ordinary Tangherlini black hole were calculated solely using the standard 6th-order WKB formula. Hence, to the best of our knowledge, we also present their accurate values for the first time.

\section{Stability of perturbations}\label{sec:stability}

A crucial inquiry related to the ensemble of quasinormal modes for a field's perturbations under consideration concerns its stability, indicated by the absence of unboundedly growing modes in the spectrum.
Since, for $r>r_0$,
$$f(r)>0, \quad f'(r)>0,$$
the effective potential for the scalar perturbations $V_0(r)$ is positive everywhere outside the black hole. Hence, a self-adjoint operator,
\begin{equation}\nonumber
-\frac{d^2}{dr_*^2}+V_0
\end{equation}
is positive definite, and, consequently, all the eigenvalues $\omega$ of Eq.~(\ref{wavelike}) have negative imaginary part (see Sec.~VIII in~\cite{Konoplya:2011qq} for details). Therefore, the scalar field perturbations are linearly stable.

Although the effective potential $V_1$ is not positive definite, but has a negative gap, we know that the electromagnetic perturbations are governed by a positive self-adjoint operator for any spherically symmetric background (see Eq.~(2.18) and~(2.21) in~\cite{Crispino:2000jx}). In particular, it is easy to see that
\begin{equation}\nonumber
-\frac{d^2}{dr_*^2}+V_1\equiv-\left(\frac{d}{dr_*}+\frac{D - 4}{2r}f(r)\right)^2+\frac{\ell(\ell+D-3)}{r^2}
\end{equation}
is also a positive self-adjoint operator.

Thus, we conclude that the test scalar and electromagnetic fields do not have unstable modes in their spectrum.

\section{Conclusions}\label{sec:conclusions}

In the present work, we have demonstrated that:
\begin{itemize}
\item The $D$-dimensional generalization of the regular Dymnikova black hole cannot be a solution in the perturbative theory with an infinite series of higher curvature corrections, unlike many other regular solutions described recently in \cite{Bueno:2024dgm}. However, this does not exclude the possibility of the Dymnikova spacetime being a solution of such a non-perturbative (in coupling) theory.
\item We have computed the quasinormal modes of scalar and electromagnetic perturbations in the background of the $D$-dimensional regular Dymnikova black hole using two independent methods: the higher-order WKB method with Padé approximants and the Bernstein polynomial method, achieving good agreement between both methods.
\item As a by-product, we have identified accurate quasinormal frequencies of electromagnetic perturbations of the $D$-dimensional Tangherlini black hole, which appear to have been omitted in the literature.
\end{itemize}

Here, our study has been limited to investigating low-lying modes and did not consider the behavior of overtones. However, it is worth noting that the metric deviates from the Schwarzschild metric primarily in the near-horizon zone and gradually merges with the Schwarzschild metric in the distant region. This suggests that overtones may exhibit much greater sensitivity to such deformations compared to the fundamental mode, possibly resulting in the outburst of the first few overtones \cite{Konoplya:2022pbc,Konoplya:2023aph}. We hope that future studies will shed light on this question.

We note that, because the metric function (\ref{HDDymnikova}) cannot be expanded in the series of $\alpha$, the accurate Frobenius method cannot be applied to the expansion of the equation to such an approximate metric. In order to use the Frobenius method for the Dymnikova black hole it was proposed in \cite{Konoplya:2023aph} to use the parametrization \cite{Rezzolla:2014mua}, which was generalized for $D>4$ in \cite{Konoplya:2020kqb}. This way the accurate values of the higher overtones could be accurately calculated.

\end{document}